\documentclass[USletter,11pt,twoside]{article}

\usepackage[utf8]{inputenc}
\usepackage[english]{babel}
\usepackage[T1]{fontenc}
\usepackage{url}
\usepackage[colorlinks=true, allcolors=blue]{hyperref}
\usepackage{graphicx}
\usepackage[left=85pt, right=85pt, top=75pt, bottom=85pt,
headheight=15pt,%
headsep=10pt]{geometry}%
\usepackage{amsmath, amssymb}
\usepackage{authblk}
\usepackage{algorithm, algpseudocode}
\usepackage{mathptmx}

\usepackage[figurename=Fig.]{caption}

\DeclareSymbolFont{matha}{OML}{txmi}{m}{it}
\DeclareMathSymbol{\varv}{\mathord}{matha}{118}

\title{Direct laser written aperiodic photonic volume elements for complex light shaping with high efficiency: inverse design and fabrication}

\author[1,2,4,*]{Nicolas Barré}
\author[3]{Ravi Shivaraman}
\author[1]{Simon Moser}
\author[3]{Patrick Salter}
\author[2,4]{Michael Schmidt}
\author[3,4]{Martin J. Booth}
\author[1,4,*]{Alexander Jesacher}

\affil[1]{Institute of Biomedical Physics, Medical University of Innsbruck, M\"{u}llerstra\ss e 44, 6020 Innsbruck, AT}
\affil[2]{Institute of Photonic Technologies, Friedrich-Alexander-University Erlangen-N\"{u}rnberg, Konrad-Zuse-Stra\ss e 3/5, Erlangen 91052, DE}
\affil[3]{Department of Engineering Science, University of Oxford, Parks Road, Oxford OX1 3PJ, UK}
\affil[4]{Erlangen Graduate School in Advanced Optical Technologies (SAOT), Friedrich-Alexander-University Erlangen-N\"{u}rnberg, Paul-Gordan-Stra\ss e 6, 91052 Erlangen, DE}

\affil[*]{Corresponding authors: nicolas.barre@protonmail.com \\ alexander.jesacher@i-med.ac.at}

\begin{document}

\maketitle

\begin{abstract}
  Light plays the central role in many applications. The key to
  unlocking its versatility lies in shaping it into the most
  appropriate form for the task at hand.  Specifically tailored
  refractive index modifications, directly manufactured inside glass
  using a short pulsed laser, enable an almost arbitrary control of
  the light flow. However, the stringent requirements for quantitative
  knowledge of these modifications, as well as for fabrication
  precision, have so far prevented the fabrication of light-efficient
  aperiodic photonic volume elements (APVEs).
  
  Here we present a powerful approach to the design and manufacturing
  of light-efficient APVEs. We optimize application-specific 3D
  arrangements of hundred thousands of microscopic voxels and
  manufacture them using femtosecond direct laser writing inside
  millimeter-sized glass volumes. We experimentally achieve
  unprecedented diffraction efficiencies up to 80\%, which is enabled
  by precise voxel characterization and adaptive optics during
  fabrication.
  
  We demonstrate APVEs with various functionalities, including a
  spatial mode converter and combined intensity shaping and
  wavelength-multiplexing.  Our elements can be freely designed and
  are efficient, compact and robust. Our approach is not limited to
  borosilicate glass, but is potentially extendable to other
  substrates, including birefringent and nonlinear materials, giving a
  preview of even broader functionalities including polarization
  modulation and dynamic elements.
\end{abstract}

\clearpage

\section{Introduction}\label{sec:Intro}

Light has become an indispensable tool in our modern day societies. It
plays the central role in countless technological solutions, such as
in the information transport along glass fibers or in various display,
imaging and sensing applications~\cite{rossing2019photonics}. The
increasing significance of light as a tool sparked a sharply rising
demand of technologies that enable the temporal and spatial sculpting
of light in user-defined ways~\cite{rubinsztein2016roadmap,
forbes2021structured, piccardo2021roadmap}.

Of notable importance are so-called multiplexing applications, where
one of many pre-defined output fields can be created upon presenting a
unique pre-defined input field. Elements that can handle such tasks
with high efficiency would find numerous applications in the analysis
and synthesis of light fields. Examples range from conventional
spectroscopy to complex tasks around information processing and
transport. Mode-division multiplexing for example is a potential
solution for avoiding the threat of reaching an upper limit in
communication speed (``capacity crunch'')~\cite{richardson2013space,
li2014space, winzer2018fiber, puttnam2021space}. Meanwhile optical
computing is currently experiencing a revival thanks to novel ways of
manufacturing computer-designed optical networks which enable
sophisticated data processing at the speed of
light~\cite{dinc2020computer, porte2021direct, moughames2020three,
luo2022computational, liao2021all}.

Optically recorded volume holograms, e.g. from photorefractive
materials~\cite{gunter2007photorefractive}, have traditionally been
used for multiplexing applications~\cite{coufal2000holographic,
wakayama2013mode}. However, creating such holograms requires the
realization of matching physical interference patterns at the
recording step, and the approach is also limited to photosensitive
substrates, which severely limits its practical applicability.

Aperiodic Photonic Volume Elements (APVE) can circumvent these
limitations: they are designed on the computer and open to a multitude
of different manufacturing techniques. For instance, they can be
directly written into a substrate by locally changing its 3D
refractive index (RI) distribution in a voxel by voxel fashion.
However, a major challenge in the production of APVEs is to meet the
high accuracy requirements for manufacturing since producing an
efficient APVE puts high demands on the ability to modify the 3D RI
distribution with high precision. Since the output light field arises
from the interference of thousands of scattered waves, even small
errors made on the voxel-scale quickly accumulate and severely degrade
the output field quality. For this reason, past demonstrations of
manufacturing APVEs in glass have been limited to proof of concept
studies with rather low diffraction
efficiencies~\cite{gerke2010aperiodic} or restricted to 2D
configurations~\cite{Douglass:22}. Recently, 2-photon
polymerization additive manufacturing of multi-layer
holograms~\cite{dinc2020computer} and the fabrication of 3D graded
index materials~\cite{porte2021direct} have been demonstrated.

Here we introduce a novel approach for realizing APVEs using direct
laser writing in transparent dielectric media such as glass. Our
method allows for obtaining high diffraction efficiencies of up to
80\%, many times more than what could previously be
achieved~\cite{gerke2010aperiodic}. This is made possible by
employing precise tomographic voxel
characterization~\cite{barre2021tomographic} in combination with a
design algorithm based on numerical beam propagation and the use of
adaptive optics to ensure space invariant voxel shapes throughout the
full body of the APVE. We experimentally demonstrate the design and
manufacture of three different, highly integrated APVEs that are
optimized for intensity shaping, spatial mode and wavelength
multiplexing.

We believe that our results represent an important
step towards the realization of robust and highly integrated 3D
light shapers for many important application fields, such
as information transport~\cite{puttnam2021space}, optical
computing~\cite{dinc2020computer, porte2021direct, moughames2020three,
luo2022computational, liao2021all}, the imaging through multimode
fibers~\cite{butaite2022build} and nonlinear
photonics~\cite{Wright:22}. They may further help to obtain
fundamental insights in the behaviour of quantum states of light upon
scattering~\cite{lib2022quantum}.

\section{APVE concept}\label{sec:Concept}

\begin{figure}[!ht]
  \centering \includegraphics[width=0.9\textwidth]{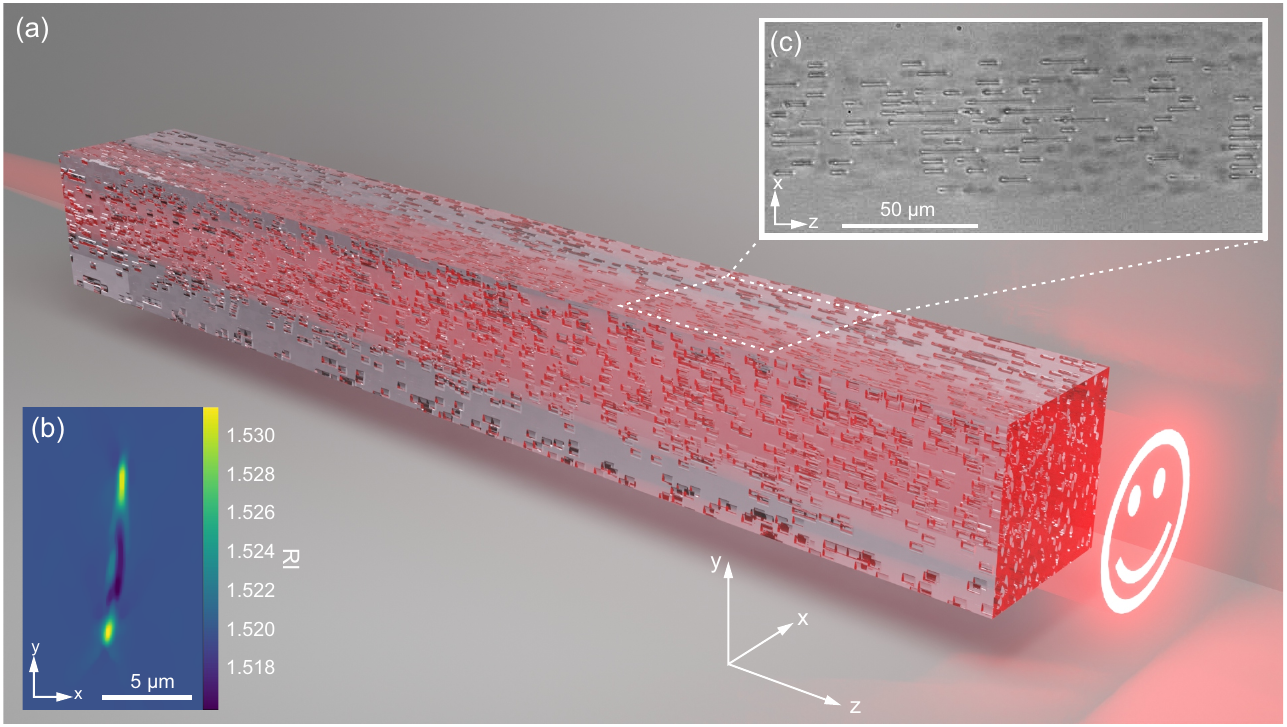}
  \caption{\textbf{Light manipulation with an aperiodic photonic volume element.}
    (a) Sketch of a laser processed glass substrate containing many
    voxels of modified RI. (b) Tomographically measured RI cross
    section of a single voxel. (c) Widefield image taken from a
    fabricated device.}
  \label{fig:concept}
\end{figure}

Figure~\ref{fig:concept}~(a) illustrates the basic concept of our
photonic volume element, which is a laser processed 3D region
inside a glass substrate, containing hundreds of thousands of
spatially separated voxels of modified refractive index at pre-defined
positions. 
Our approach is based on past demonstrations of direct laser written aperiodic volume optics~\cite{gerke2010aperiodic}, but also inspired by previous work on computer-generated 2D holograms based on waveguides~\cite{berlich2016fabrication}.

A single voxel measures only
1.75$\times$7.5$\times$10~\textmu m$^3$ and is created by moving the
z-stage (the long horizontal axis) for 10~\textmu m during exposure to
the laser focus. Fabrication of the whole device takes about 20
minutes. The quantitative RI profile of a single voxel across the
x-y-axes, fabricated in Eagle glass, is shown in (b). This profile
must be precisely known, as it serves as the basis for the numerical
design process. The RI profile characterization is performed in a
home-build tomographic microscope~\cite{barre2021tomographic}. The RI
distribution of each voxel is assumed to be invariant along the
z-direction.  The input and output fields of the APVE are coupled
through its smallest end facets. A widefield transmission image of a
fabricated APVE is shown in (c).

Our design algorithm operates on a Cartesian grid of size
$I\times~J\times~K$, following the directions $(x,y,z)$ of the
coordinate system sketched in (a), which defines potential voxel
positions. For our demonstrators, $I=55$, $J=14$ and $200 \leq K \leq
400$ depending on the device extension along the propagation direction
$z$. This corresponds to a maximum number of voxels varying between
154,000 and 308,000. The number of voxels in the transverse $x$- and
$y$-directions are quite different due to the elongated nature of our
reference voxel represented in (b). Basically, the algorithm
simulates the light flow through the device using a beam propagation
method~\cite{feit1978light} and decides upon the activation of a
voxel, depending on whether it would improve the output wavefront or
not. For our experimental demonstration, we explored only binary
designs made of identical voxels depicted in (b), but the inverse
design algorithm we implemented generalizes well to multilevel RI
modifications. Details about this algorithm are provided in the
Appendix~\ref{sec:Design_algorithm}.

The successful implementation of our APVE requires not only precise
knowledge about the voxel properties, but also the manufacturing
abilities to produce the same profile reliably at all targeted
substrate depths. This can be ensured using dynamic wavefront control,
which compensates for spherical aberrations arising when focusing the
laser into the substrate~\cite{jesacher2010parallel}. Further
information about the manufacturing is also provided in
Appendix~\ref{sec:Manufacturing_APVEs}.

\section{Results}\label{sec:results}

In the following, we explore three types of APVEs, designed for specific
applications with different levels of complexity. The first one is a
single-mode intensity shaper with a length of only 2~mm and discussed
in section~\ref{sec:intensity}, the second one is a multi-colour (RGB)
multiplexer of 3~mm length and discussed in
section~\ref{sec:RGB}. Finally, a Hermite--Gaussian 6-mode sorter of
4~mm length is discussed in section~\ref{sec:mode-mulitplexing}.

\subsection{Sculpting intensity distributions}\label{sec:intensity}

Our APVEs can be optimized to shape user-defined intensity patterns
with high efficiency. To demonstrate this, we designed and fabricated
an element producing a microscopic ``smiley'' at its output facet
(i.e. the last layer of voxels) when read out with a Gaussian beam at
zero degree readout angle (vacuum wavelength $\lambda_0$ = 640~nm,
waist $w_0$ = 40~\textmu m). The physical dimensions of the element are
0.1$\times$0.1$\times$2~mm$^3$ and it contains
55~$\times$~14~$\times$~200 voxels. As mentioned earlier, the unequal
grid sizes along the $x$- and $y$-directions are due to the
anisotropic voxel shape, which is a consequence of the likewise
anisotropic shape of the focus creating the voxel. The device itself,
i.e., the 3D region containing fabricated voxels, is embedded inside a
larger glass chip for easier handling.

\begin{figure}[h]
  \centering \includegraphics[width = 12cm]{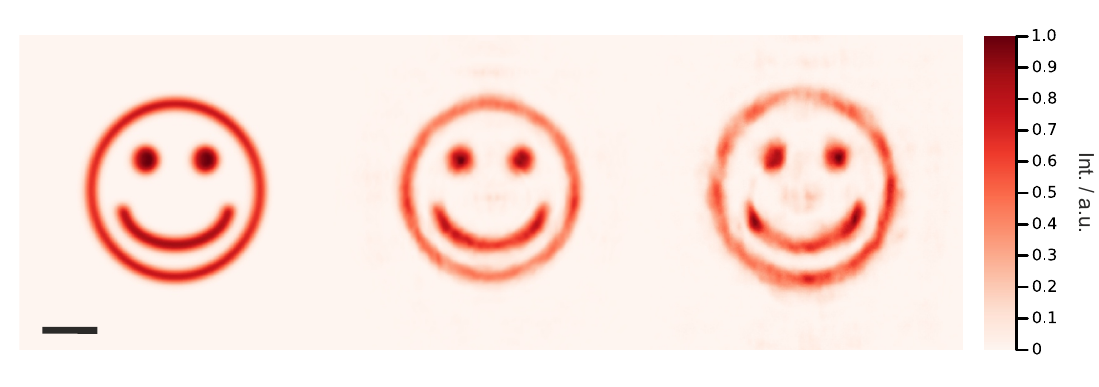}
  \caption{\textbf{Results from a ``smiley'' generator.} Left:
    Designed output intensity; Middle: simulated readout; right:
    experimental result. The total light efficiency $\eta_{tot}$ of
    the experimental result is about 80\%. The scale bar measures
    20~\textmu m.}
  \label{fig:smiley}
\end{figure}

Figure~\ref{fig:smiley} provides results obtained with this APVE.
It shows the designed output intensity template for a Gaussian input
of 40~\textmu m waist as well as the simulated and experimentally
obtained results. The simulated result was obtained by numerical
propagation of the input beam through a computer model of the
optimized voxel arrangement. It thus marks an upper limit for the
quality achievable in the experiment. The experimental result was
obtained by imaging the output plane with a regular widefield
microscope (see Appendix~\ref{sec:APVE_carac} for additional information
about the experimental APVE characterization).

The diffraction efficiency of the APVE has been determined by
calculating an overlap integral of the target amplitude pattern $u_t$
with the simulated and experimentally measured output amplitude
patterns $u_{out}$ over the area of the observable end facet $A$:

\begin{equation}
    \eta = \left|\int_A u_{t}~u_{out}^*~dA\right|^2,
\label{eq:eta}
\end{equation}
where the powers of $u_{t}$ and $u_{out}$ are both normalized to 1.
We introduce another efficiency metric $\eta_{tot}$, which further
takes into account scattering losses into non-observable angles:
\begin{equation}
    \eta_{tot} = T~\eta,
\label{eq:eta_tot}
\end{equation}
with transmission factor $T \leq 1$. Therefore, $\eta$ quantifies the
power fraction of the measurable \emph{output} light\footnote{the
  light arriving at the (simulated or real) camera plane} which is
effectively turned into the target distribution, while $\eta_{tot}$ is
the respective conversion efficiency for the \emph{input} beam. Losses
due to Fresnel reflections at the input/output facets of the APVE are
neglected in this consideration.

The theoretical and experimental diffraction efficiencies $\eta_{tot}$
of the smiley converter are 90\% and 80\%, respectively. We note that
$T\approx$1 for the smiley shaper in both simulation and experiment,
i.e., the amount of scattering losses are close to zero. Since this
application aims at intensity-only shaping, the phase distributions of
$u_{t}$ and $u_{out}$ are irrelevant and thus assumed to be flat in
Eq.~\ref{eq:eta}. We further derived an error metric $\epsilon$,
defined as the root mean square (RMS) difference between normalized
target and simulated/experimental output intensities. We find
$\epsilon_{\mathrm{sim}} = 0.25, ~\epsilon_{\mathrm{exp}} = 0.35$.

Therefore, even though the intensity shaping shown here can be easily
achieved using 2D diffractive optical elements, we are nevertheless
able to simply demonstrate that our computational design approach can
be faithfully implemented through the experimental characterization
and reproducible manufacture of 3D voxel arrangements inside glass.

\subsection{Wavelength multiplexing}\label{sec:RGB}
\begin{figure}[!ht]
  \centering
  \includegraphics[width=0.8\textwidth]{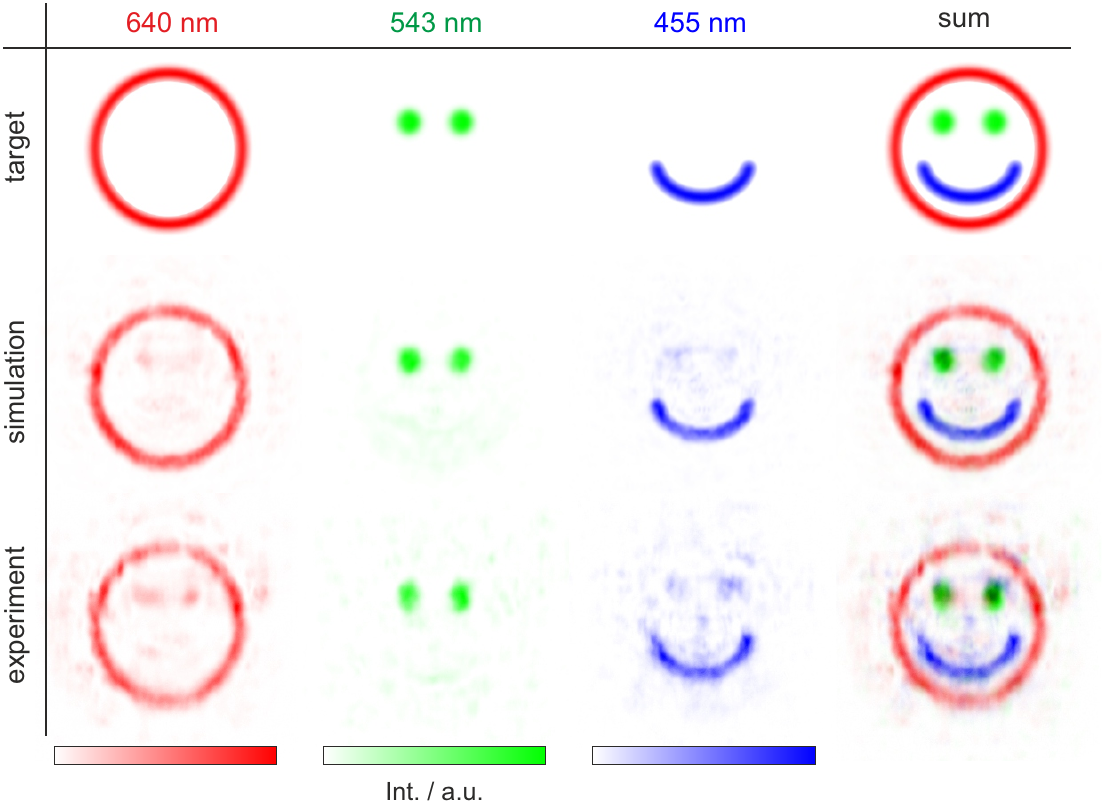}
  \caption{\textbf{Wavelength multiplexing.} Different parts of the
    smiley appear depending on the readout wavelength. Top row: target
    intensity patterns used for the APVE design. Middle row: results
    from a simulated readout. Bottom row: experimental readouts.}
  \label{fig:colour smiley}
\end{figure}

We demonstrate the feasibility of colour multiplexing by designing and
fabricating a multi-colour ``smiley'' shaper, where different parts of
the smiley (eyes, mouth, head) appear only for specific readout
wavelengths as illustrated in Fig.~\ref{fig:colour smiley}. This APVE
measures 0.1$\times$0.1$\times$3~mm$^3$ and contains
55~$\times$~14~$\times$~300 voxels. When read out at 640~nm with a
beam waist of $w_0$ = 40~\textmu m, only the circular head appears at
the output. Similarly, the eyes appear for a readout wavelength of
543~nm ($w_0$ = 30~\textmu m) and the mouth for 455~nm ($w_0$ =
20~\textmu m). Results obtained from simulated readouts and
experiments using fiber-coupled lasers (see Appendix~\ref{sec:Additional_color_multiplexer} for details) 
are summarized in Fig.~\ref{fig:colour smiley}. 
The image table depicts false-colour
intensity distributions at the output facet of the APVE, which are
individually normalized to their respective peak intensities. The
fourth image column on the right shows a computer-generated overlap of
the three readouts.

The simulated and experimentally determined power conversion
efficiencies $\eta$ from the three Gaussian input beams to the
respective facial structures are summarized in table~\ref{tab:RGB}.
The table values are calculated using Eq.~\ref{eq:eta}, i.e., they
show which percentage of the \emph{output} laser power is shaped into
the respective facial structures. The average simulated efficiency
over all colours is 67\% for the simulation and 55\% in the
experiment. From the table values it is noticeable that in particular
the blue wavelength shows the largest disparity between simulation and
experiment, with an efficiency of 64\% in the simulation and 49\% in
the experiment. This may be attributed to small remaining differences
between the assumed and real voxel profiles at length scales of a few
hundred nanometres. Such small irregularities will especially affect
shorter wavelengths and are likely to be overlooked by our tomographic
voxel characterization method, which employs the same wavelength for
inspection and a smoothing total variation norm
regularizer~\cite{barre2021tomographic}. The simulated and
experimentally measured transmission values $T$ for all three readout
wavelengths are shown in table~\ref{tab:T}. We observe well matching
numbers between simulation and experiment for the red and green
wavelengths, while the blue transmission value is around 25\% lower in
the experiment than predicted by the simulation.

\begin{table}[!ht]
\begin{center}
\begin{tabular}{ |c|c|c|c| } 
  \hline
  $\lambda_0 / \textrm{nm}$ & head & eyes & mouth\\
  \hline 
  640 & \textbf{77} & 2  & 1\\ 
  543 & 8 & \textbf{60} & 7 \\ 
  455 & 3 & 7 & \textbf{64}\\ 
  \hline
\end{tabular}
\quad
\begin{tabular}{ |c|c|c|c| } 
  \hline
  $\lambda_0 / \textrm{nm}$ & head & eyes & mouth\\
  \hline 
 640 & \textbf{65} & 7 & 6\\ 
 543 & 5 & \textbf{50} & 6\\ 
 455 & 1 & 4 & \textbf{49}\\ 
 \hline
\end{tabular}
\end{center}
\caption{\textbf{Table of simulated (left) and experimentally (right)
    obtained conversion efficiencies $\eta$.} For each wavelength, the
  numbers state the respective percentage of the output power forming
  head, eyes and mouth.}
\label{tab:RGB}
\end{table}

\begin{table}[!ht]
\begin{center}
  \begin{tabular}{ |c|c|c|c| } 
    \hline
    & 640 nm & 543 nm & 455 nm\\
    & (head) & (eyes) & (mouth)\\
    \hline 
    T (sim) & 83   & 82   & 87  \\ 
    T (exp) &  84   & 82   & 66 \\ 
    \hline
  \end{tabular}
\end{center}
  \caption{\textbf{Table of simulated and experimentally measured
      transmission factors $T$ for the colour smiley APVE.}}
  \label{tab:T}
\end{table}

The simulated (experimental) total efficiencies $\eta_{tot}$ for the
head, eyes and mouth features are therefore 64\% (55\%), 49 (41\%) and
56\% (32\%).

We further investigated the wavelength-dependent APVE properties using
a fiber-coupled mono\-chromator (Polychrome IV from TILL Photonics),
whose output wavelength was tuned from 420~nm to 680~nm in steps of
10~nm. In contrast to the laser readouts where each beam had an
individual, optimal waist value, the beam waist of the monochromator
light at the APVE input facet was about 30~\textmu m for all
wavelengths (details are described in Appendix~\ref{sec:Additional_color_multiplexer}).
While this means that the monochromator results are expected to be
somewhat sub-optimal they nevertheless provide valuable information
about the wavelength dependence of the APVE.

\begin{figure}[ht]
  \centering
  \includegraphics[width=0.8\textwidth]{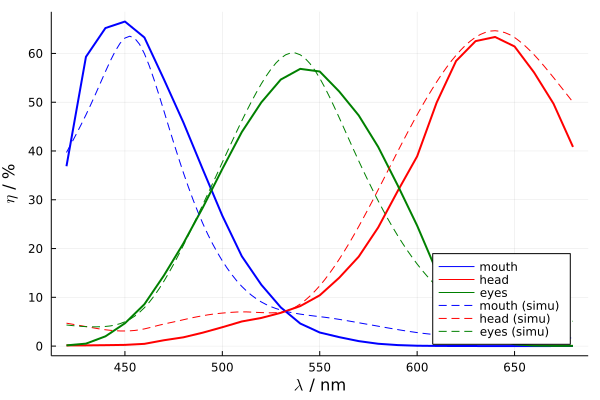}
  \caption{\textbf{Power conversion efficiencies of the multi-colour
      APVE.} The solid curves indicate the measured percentage of the
    output power transformed into the features "mouth", "eyes" and
    "head" depending on the readout wavelength. The dashed lines
    correspond to simulated readouts.}
  \label{fig:RGB monochromator}
\end{figure}

At each wavelength we took images of the input beam and the APVE
output (see data in Appendix~\ref{sec:Additional_color_multiplexer}), which allowed us to
calculate wavelength dependent power conversion efficiencies according
to Eq.~\ref{eq:eta} for each spatial feature (mouth, eyes, head).
These efficiencies are plotted in Fig.~\ref{fig:RGB monochromator}. Of
note, the peaks appear at wavelengths which almost exactly match the
targets at 455~nm, 543~nm and 640~nm. The measured efficiencies $\eta$
at these wavelengths are 63\%, 57\% and 63\% and thus close to the
simulation values obtained from the laser readouts as shown in
table~\ref{tab:RGB} (left). Interestingly, the blue monochromator
image of the "mouth" shows a significantly higher efficiency than the
respective experimental blue laser readout (63\% vs. 49\%).

Figure~\ref{fig:RGB monochromator} further contains data from
simulated readouts (dashed lines), which resemble the experimental
data. To match the experimental conditions, these simulations
assumed a beam waist of 30~\textmu m for all readout wavelengths,
which cause the peak efficiency values to be slightly lower than the
values in table~\ref{tab:RGB}, which assume readouts taken at the
designed waist values.

\subsection{Angular multiplexing}\label{sec:mode-mulitplexing}
The high angular selectivity of a photonic volume element allows for encoding
different outputs for varying input angles. This characteristic makes
APVEs highly interesting for tasks such as mode division multiplexing
and sorting. To investigate the feasibility of angular multiplexing
for our approach, we have designed a APVE measuring only
0.1$\times$0.1$\times$4~mm$^3$ and containing
55~$\times$~14~$\times$~400 voxels, which produces different
Hermite-Gaussian (HG) modes depending on the angle of incidence (AOI)
of a Gaussian readout beam.

Such a device could be used as a mode division multiplexer to enhance
data transfer speed as sketched in Fig.~\ref{fig:mode sorting
  principle}: multiple signals are delivered by single mode fibers in
a triangular arrangement, which has been shown to facilitate high
Hermite-Gaussian (HG) mode conversion
efficiencies~\cite{fontaine2019laguerre}. The fiber ends are in the
focal plane of a convex lens, which collimates the light leaving the
fiber outputs and sends it into the APVE at different, fiber-specific
propagation angles. The APVE converts each signal into a particular HG
mode travelling in parallel to the optical axis, such that it can be
efficiently coupled into a single multimode fiber.
\begin{figure}[ht]
  \centering \includegraphics[width=0.8\textwidth]{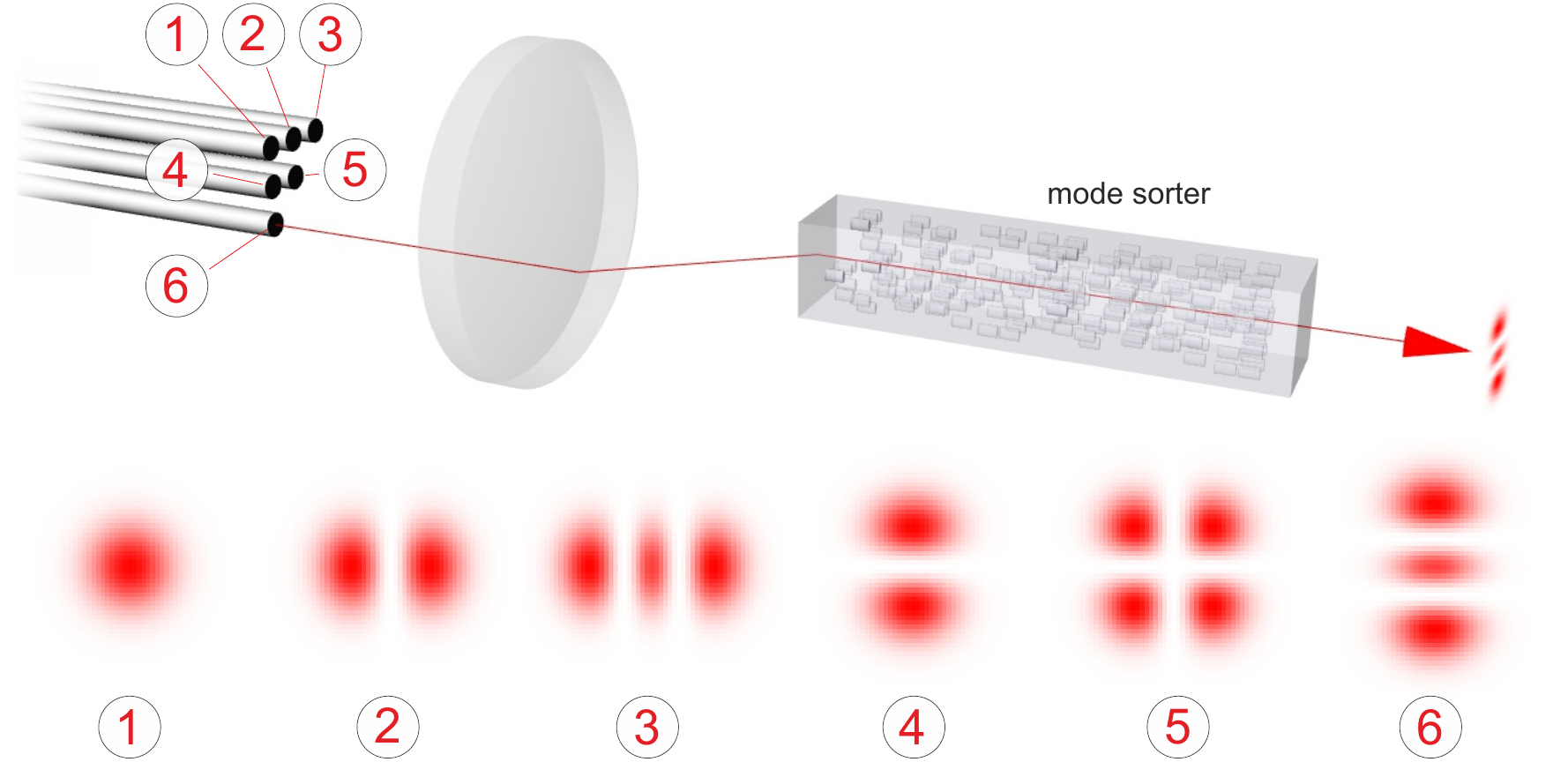}
  \caption{\textbf{Principle of mode division multiplexing with our
      mode sorter.} Multiple signals are delivered via single mode
    fibers, arranged in a triangle. A lens gives each input beam a
    specific AOI. The APVE transforms each input beam into one of six
    different HG modes.}
  \label{fig:mode sorting principle}
\end{figure}

We designed a mode sorting APVE for readout with a Gaussian input of
640~nm wavelength and 25~\textmu m waist.  Altering the input beam's
AOI by merely 1.4 degrees (defined in air) transforms the output field
from one particular Hermite-Gaussian (HG) mode to another.  In total
we encode six different HG modes as shown in Fig.~\ref{fig:mode
  sorting principle}.

\begin{figure}[ht]
  \centering \includegraphics[width=0.8\textwidth]{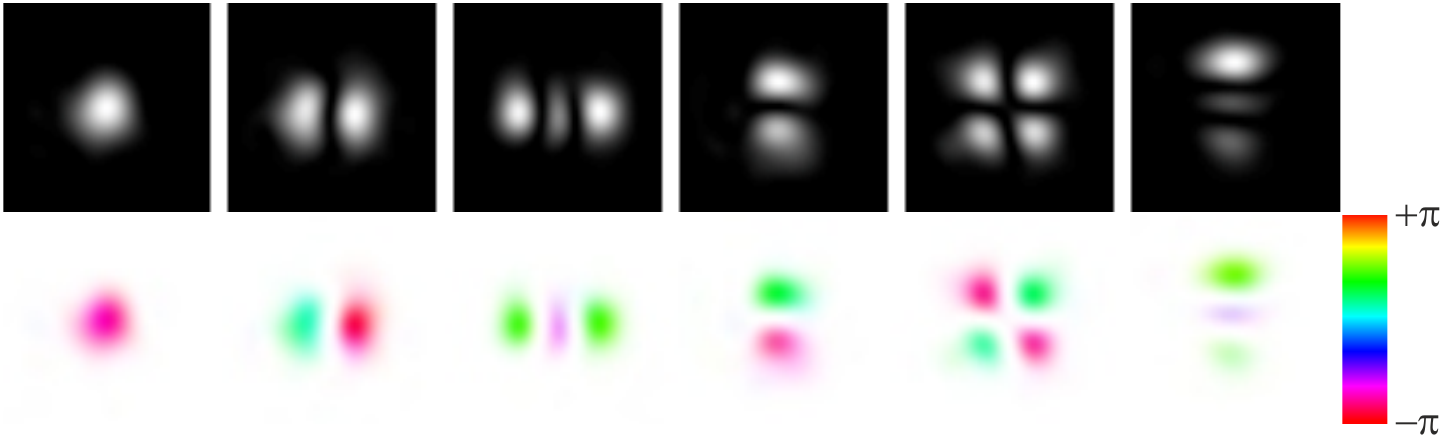}
  \caption{\textbf{Simulated results from the mode sorter.} The
    images show intensities (top row) and phases when reading out with
    a Gaussian beam ($w_0$ = 25~\textmu m, $\lambda_0$ = 640~nm) at
    six different AOIs. Each angle produces a different HG mode at the
    output. The saturation of the phase images is weighted by the
    intensity for enhanced clarity.}
  \label{fig:mode sorter_simu}
\end{figure}

Figure~\ref{fig:mode sorter_simu} shows the output intensities and
phases resulting from a simulated readout of the mode sorting
APVE. The matching experimental data are shown in Fig.~\ref{fig:mode
  sorter_exp}. For both, simulation and experiment, the numerical aperture (NA) at the
output needs to be reduced in order to block stray light produced by
the element. This reduces the overall light efficiency but enhances
the quality of the generated modes. In simulation the NA was
restricted to 0.02, in experiment to about 0.05.

\begin{figure}[ht]
  \centering \includegraphics[width=0.8\textwidth]{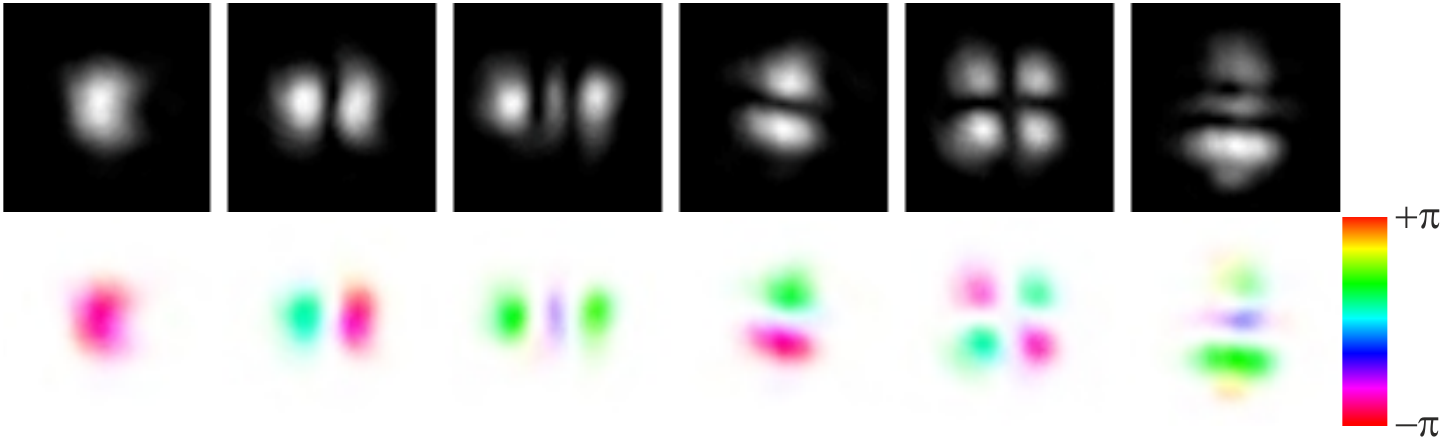}
  \caption{\textbf{Experimental results from the mode sorter.}
    The images show intensities (top row) and phases when reading out
    the APVE with a Gaussian beam ($w_0$ = 25~\textmu m, $\lambda_0$ =
    640~nm) at six different incidence angles.}
  \label{fig:mode sorter_exp}
\end{figure}

As for the APVEs discussed previously, the overall quality of the
mode sorter is governed by two factors: (i) the purity of the
generated modes and (ii) the transmission of the APVE quantified
by parameter $T$. Although the element itself absorbs almost no
light, there is some power loss caused by the aperture stop
restricting the imaging NA. We quantified point (i) by calculating
efficiency values in analogy to Eq.~\ref{eq:eta} and taking the
different HG modes as target fields:

\begin{equation}
  \eta_{i,j} = \left|\int_A \mathrm{HG}_{i,j}~u_{out}^*~dA\right|^2,
\label{eq:OI}
\end{equation}

where $\mathrm{HG}_{i,j}$ and $u_{out}^*$ represent the target mode
and conjugate output field with their total powers normalized to 1.
An efficiency value of $\eta_{i,j} = 1$ means that the field has been
shaped into a perfect $\mathrm{HG}_{i,j}$ mode, whereas
$\eta_{i,j} = 0$ means that the field is orthogonal to this mode.
Tables~\ref{tab:OI_simu} and \ref{tab:OI_exp} contain efficiency
values of the simulated and experimental mode sorter. The off-diagonal
terms represent undesired crosstalks into neighbouring channels. The
purest modes have efficiency values of 90\% in the simulation and 88\%
in the experiment. The highest crosstalk values are 3\% for the
simulation and 5\% for the experiment.

The mode-dependent transmission of the APVE (ii), including both
losses due to scattering and spatial filtering by the restricted
imaging NA, are summarized in table~\ref{tab:transmissions
  sorter}. These results are still far from what could be
theoretically obtained with a gradient index
design~\cite{barre2022inverse}, but they are surprisingly good for a
binary design. Introducing different types of voxels as new degrees of
freedom for the inverse design optimization could significantly
improve the capabilities of these mode multiplexers and approach the
maximum performance of gradient index designs, for which a mature
manufacturing technology is not yet available in glass.

\begin{table}[ht]
\begin{center}
\begin{tabular}{ |c|c|c|c|c|c|c| } 
 \hline
 input angle no. & HG$_{00}$ & HG$_{10}$ & HG$_{20}$ & HG$_{01}$ & HG$_{11}$ & HG$_{02}$\\
 \hline 
 1 & $\textbf{90.6}$ & $0.9$ & $0.6$ & $2.0$ & $0.1$ & $0.3$\\
 2 & $0.2$ & $\textbf{89.1}$ & $0.9$ & $0.3$ & $3.4$ & $0.1$\\
 3 & $0.9$ & $0.5$ & $\textbf{90.2}$ & $0.4$ & $0.2$ & $0.1$\\
 4 & $0.9$ & $1.1$ & $0.2$ & $\textbf{87.2}$ & $0.1$ & $0.4$\\
 5 & $0.1$ & $1.0$ & $0.0$ & $0.2$ & $\textbf{93.9}$ & $0.2$\\
 6 & $1.1$ & $0.1$ & $0.1$ & $1.6$ & $0.1$ & $\textbf{79.6}$\\
 \hline
\end{tabular}
\end{center}
\caption{\textbf{Table of simulated efficiency values $\eta_{i,j}$ in
    percent.} For each input angle, the numbers state the respective
  power fractions (in percent) of the transmitted output light that is
  shaped into the corresponding HG modes.}
\label{tab:OI_simu}
\end{table}

\begin{table}[ht]
\begin{center}
\begin{tabular}{ |c|c|c|c|c|c|c| } 
 \hline
 input angle no. & HG$_{00}$ & HG$_{10}$ & HG$_{20}$ & HG$_{01}$ & HG$_{11}$ & HG$_{02}$\\
 \hline 
 1 & $\textbf{88.4}$ & $1.1$ & $2.6$ & $0.3$ & $5.0$ & $3.5$\\ 
 2 & $0.3$ & $\textbf{87.0}$ & $0.3$ & $1.8$ & $0.7$ & $2.5$\\ 
 3 & $0.7$ & $0.5$ & $\textbf{83.8}$ & $0.1$ & $0.1$ & $0.2$\\  
 4 & $0.6$ & $1.6$ & $0.1$ & $\textbf{83.7}$ & $0.4$ & $4.4$\\  
 5 & $2.3$ & $1.4$ & $0.7$ & $0.6$ & $\textbf{86.6}$ & $1.2$\\ 
 6 & $1.0$ & $1.4$ & $1.2$ & $4.5$ & $1.3$ & $\textbf{65.7}$\\  
 \hline
\end{tabular}
\end{center}
\caption{\textbf{Table of experimental efficiency values $\eta_{i,j}$
    in percent.}}
\label{tab:OI_exp}
\end{table}

\begin{table}[ht]
\begin{center}
\begin{tabular}{ |c|c|c|c|c|c|c| } 
 \hline
  & HG$_{00}$ & HG$_{10}$ & HG$_{20}$ & HG$_{01}$ & HG$_{11}$ & HG$_{02}$\\
 \hline 
 T (sim) & $50.0$ & $49.2$ & $49.5$ & $42.2$ & $57.9$ & $48.9$\\ 
 T (exp) & $21.6$ & $29.6$ & $20.4$ & $38.8$ & $33.0$ & $32.3$\\ 
 \hline
\end{tabular}
\end{center}
\caption{\textbf{Table of simulated and experimental transmission
    values $T$.} For each mode, the numbers state the output/input
  power ratio in percent.}
\label{tab:transmissions sorter}
\end{table}

\clearpage

\section{Discussion}\label{sec:Discussion}
In this discussion section, we place our method in the context of
related technological approaches.

\subsection*{Optically recorded volume holograms} An important
difference between traditional volume holograms and our APVEs exists in
the physics of their creation. Traditional holograms are optically
recorded by transferring the properties of interference patterns into
a photosensitive material, such as a
photopolymer~\cite{colburn1971volume, jialing2019review} or a
photorefractive crystal~\cite{hesselink1998photorefractive}, which
comes naturally with limitations. Firstly, each desired function must
be physically realized at the recording step, which becomes unfeasible
for advanced multiplexing applications. Secondly, one is bound to a
limited number of suitable photosensitive materials. Finally, the
dynamic range of optically recorded holograms is quickly consumed by
only a few multiplexed recordings, which has an adverse effect on the
obtainable diffraction efficiency~\cite{porte2021direct}.

\subsection*{Meta-holograms}
In contrast to the APVE concept presented here, meta-holograms rely on
nanostructured metallic or dielectric surfaces, on which
``meta-atoms'' interact with light in pre-defined
ways~\cite{yu2014flat, walther2012spatial, scheuer2017metasurfaces,
  dorrah2022tunable}. While the concept has proven to be especially
powerful for wavelength and polarization-multiplexing tasks, angular
multiplexing, i.e., the generation of different output fields at
varying input angles, appears harder to achieve: only up to four
multiplexed fields could be generated upon a large angular variation
of more then 60 deg.~\cite{jang2021independent}. Some folded
multiplane designs using metasurfaces have been realized recently in
order to increase the multiplexing capabilities of these
meta-holograms~\cite{oh2022adjoint}. Yet, the advantages of using
metasurfaces in this context seems limited since 3 reflections are
required to achieve only a 3-mode sorter, which is not significantly
better than what more conventional cascaded diffractive optical
elements can provide~\cite{dinc2020computer}.

\subsection*{Silicon Photonics}
The field of silicon photonics~\cite{jalali2006silicon,
  siew2021review}, i.e., the manufacturing and use of
photonic integrated circuits (PICs) inside silicon,
is attracting considerable attention since more than a decade. The
fabrication of photonic chips can rely on existing infrastructure and
lithographic techniques used in the electronic semiconductor industry
and enables structure sizes down to a few nanometres. The field of
silicon photonics is comparably mature, the demonstration of
wavelength multiplexing for instance dates back to the
1990s~\cite{trinh1997silicon} and has been further developed towards
high integration by employing modern tools of inverse
design~\cite{piggott2015inverse}. Spatial multimode operations have
been developed later, but are likewise possible in a highly integrated
fashion~\cite{li2019multimode}. Compared to silicon photonics, which
is bound to infrared wavelengths, our approach offers a wavelength
window reaching down to the UV, depending on the substrate, and is
potentially applicable to a larger variety of materials offering
different optical properties, including electro-optical and nonlinear
ones. Furthermore, glasses as substrate facilitate novel applications
linking optical and microfluidic
technologies~\cite{minzioni2017roadmap} and allow for easier light
coupling compared to silicon, where mode mismatch can cause important
insertion losses~\cite{photonic_packaging}. Lastly, the manufacturing
of 3D devices is more innate to direct laser writing than it is to
standard lithographic techniques~\cite{Yoo2016} and
puts also lower demands on the required infrastructure (no clean room
required).

\subsection*{Writing voxels for data storage}
From the manufacturing point of view, our approach is related to
storing data by writing of voxels inside polymers or
glass~\cite{strickler1991three, glezer1997three}. This application has
already reached an impressive maturity level, employing birefringent
multi-level voxels~\cite{zhang2014seemingly}. However, in contrast to
our application, data storage applications put
significantly lower demands on the knowledge about the quantitative 3D
properties of a single voxel, as well as their reproducible and
spatially invariant fabrication, since they can be read out
sequentially and any slight deviation from the assumed voxel shape
would be irrelevant since only a few Bits are encoded in a single
voxel. Conversely, hundreds of thousands voxels act jointly on the
readout light in our APVEs, such that even small systematic errors at
the single voxel level would have pronounced adverse effects on the
sculpted fields.

\section{Conclusion and summary}\label{sec:Conclusion}
We present a novel design and fabrication concept for robust and
light efficient aperiodic photonic volume elements. Our design algorithm
uses a mode-matching method and numerical beam propagation. The
manufacturing is based on directly writing an optimized 3D arrangement
of voxels into the volume of a transparent dielectric such as glass
using a focused femtosecond laser. Each voxel consists of a small
volume (approx. 1.75$\times$7.5$\times$10~\textmu m$^3$) of slightly
increased RI. Precise knowledge of the properties of a single voxel,
in conjunction with an advanced fabrication technique, allowed us to
physically realize APVEs of high efficiencies up to 80\%.

We experimentally realize three different APVE designs
demonstrating different functionalities: An intensity shaper turning a
Gaussian beam into a smiley-shaped light distribution, a wavelength
multiplexer producing different parts of the smiley for three
different readout wavelengths and a spatial mode multiplexer, which
converts a Gaussian input beam into particular HG modes depending on
the angle of incidence.

While our proof-of-concept studies already show unprecedented
performance, our APVE concept still has room for improvement in
several respects, promising significantly higher efficiency and more
complex functionalities in the future.  For instance, our APVEs are
made from a single voxel type, i.e., they are binary devices. However,
it is straightforward to extend the design concept to non-binary
devices. At the manufacturing side the use of varying laser powers,
focus shapes, or multi-pass writing could be used to fabricate
multiple different RI profiles. Non-binary devices will be able to
store more information and exhibit higher efficiencies. In addition,
our concept is extensible to different types of substrates, including
birefringent, electro-optic and nonlinear substrates, potentially
offering the possibility of polarization shaping, dynamic or even
erasable ferroelectric APVEs~\cite{xu2022femtosecond}.

\section*{Acknowledgments}
We thank Lisa Ackermann and Clemens Roider for valuable contributions
in the form of scientific discussions as well as for post processing
the APVE samples.

\section*{Funding}
The work is funded by the FWF (I3984-N36), the EPSRC UK (R004803/01)
and the DFG (409765270).

\clearpage

\let\OLDthebibliography\thebibliography
\renewcommand\thebibliography[1]{
  \OLDthebibliography{#1}
  \setlength{\parskip}{1ex}
  \setlength{\itemsep}{0pt plus 1ex}
}

\bibliographystyle{unsrt}
\bibliography{manuscript}

\begin{thebibliography}{10}

\bibitem{rossing2019photonics}
Thomas~D. Rossing and Christopher~J. Chiaverina.
\newblock {\em Photonics---Light in the Twenty-First Century}, pages 333--355.
\newblock Springer International Publishing, Cham, 2019.

\bibitem{rubinsztein2016roadmap}
Halina Rubinsztein-Dunlop, Andrew Forbes, M~V Berry, M~R Dennis, David~L
  Andrews, Masud Mansuripur, Cornelia Denz, Christina Alpmann, Peter Banzer,
  Thomas Bauer, Ebrahim Karimi, Lorenzo Marrucci, Miles Padgett, Monika
  Ritsch-Marte, Natalia~M Litchinitser, Nicholas~P Bigelow,
  C~Rosales-Guzm{\'{a}}n, A~Belmonte, J~P Torres, Tyler~W Neely, Mark Baker,
  Reuven Gordon, Alexander~B Stilgoe, Jacquiline Romero, Andrew~G White, Robert
  Fickler, Alan~E Willner, Guodong Xie, Benjamin McMorran, and Andrew~M Weiner.
\newblock Roadmap on structured light.
\newblock {\em Journal of Optics}, 19(1):013001, 2016.

\bibitem{forbes2021structured}
Andrew Forbes, Michael de~Oliveira, and Mark~R Dennis.
\newblock Structured light.
\newblock {\em Nature Photonics}, 15(4):253--262, 2021.

\bibitem{piccardo2021roadmap}
Marco Piccardo, Vincent Ginis, Andrew Forbes, Simon Mahler, Asher~A Friesem,
  Nir Davidson, Haoran Ren, Ahmed~H Dorrah, Federico Capasso, Firehun~T Dullo,
  Balpreet~S Ahluwalia, Antonio Ambrosio, Sylvain Gigan, Nicolas Treps, Markus
  Hiekkamäki, Robert Fickler, Michael Kues, David Moss, Roberto Morandotti,
  Johann Riemensberger, Tobias~J Kippenberg, J{\'{e}}r{\^{o}}me Faist, Giacomo
  Scalari, Nathalie Picqu{\'{e}}, Theodor~W Hänsch, Giulio Cerullo, Cristian
  Manzoni, Luigi~A Lugiato, Massimo Brambilla, Lorenzo Columbo, Alessandra
  Gatti, Franco Prati, Abbas Shiri, Ayman~F Abouraddy, Andrea Al{\`{u}},
  Emanuele Galiffi, J~B Pendry, and Paloma~A Huidobro.
\newblock Roadmap on multimode light shaping.
\newblock {\em Journal of Optics}, 24(1):013001, 2021.

\bibitem{richardson2013space}
David~J Richardson, John~M Fini, and Lynn~E Nelson.
\newblock Space-division multiplexing in optical fibres.
\newblock {\em Nature Photonics}, 7(5):354--362, 2013.

\bibitem{li2014space}
Guifang Li, Neng Bai, Ningbo Zhao, and Cen Xia.
\newblock Space-division multiplexing: the next frontier in optical
  communication.
\newblock {\em Advances in Optics and Photonics}, 6(4):413--487, 2014.

\bibitem{winzer2018fiber}
Peter~J Winzer, David~T Neilson, and Andrew~R Chraplyvy.
\newblock Fiber-optic transmission and networking: the previous 20 and the next
  20 years.
\newblock {\em Optics Express}, 26(18):24190--24239, 2018.

\bibitem{puttnam2021space}
Benjamin~J Puttnam, Georg Rademacher, and Ruben~S Lu{\'\i}s.
\newblock Space-division multiplexing for optical fiber communications.
\newblock {\em Optica}, 8(9):1186--1203, 2021.

\bibitem{dinc2020computer}
Niyazi~Ulas Dinc, Joowon Lim, Eirini Kakkava, Christophe Moser, and Demetri
  Psaltis.
\newblock Computer generated optical volume elements by additive manufacturing.
\newblock {\em Nanophotonics}, 9(13):4173--4181, 2020.

\bibitem{porte2021direct}
Xavier Porte, Niyazi~Ulas Dinc, Johnny Moughames, Giulia Panusa, Caroline
  Juliano, Muamer Kadic, Christophe Moser, Daniel Brunner, and Demetri Psaltis.
\newblock Direct (3+ 1) d laser writing of graded-index optical elements.
\newblock {\em Optica}, 8(10):1281--1287, 2021.

\bibitem{moughames2020three}
Johnny Moughames, Xavier Porte, Michael Thiel, Gwenn Ulliac, Laurent Larger,
  Maxime Jacquot, Muamer Kadic, and Daniel Brunner.
\newblock Three-dimensional waveguide interconnects for scalable integration of
  photonic neural networks.
\newblock {\em Optica}, 7(6):640--646, 2020.

\bibitem{luo2022computational}
Yi~Luo, Yifan Zhao, Jingxi Li, Ege {\c{C}}etinta{\c{s}}, Yair Rivenson, Mona
  Jarrahi, and Aydogan Ozcan.
\newblock Computational imaging without a computer: seeing through random
  diffusers at the speed of light.
\newblock {\em eLight}, 2(1):1--16, 2022.

\bibitem{liao2021all}
Kun Liao, Ye~Chen, Zhongcheng Yu, Xiaoyong Hu, Xingyuan Wang, Cuicui Lu,
  Hongtao Lin, Qingyang Du, Juejun Hu, and Qihuang Gong.
\newblock All-optical computing based on convolutional neural networks.
\newblock {\em Opto-Electronic Advances}, 4(11):200060, 2021.

\bibitem{gunter2007photorefractive}
Peter G{\"u}nter and Jean-Pierre Huignard.
\newblock {\em Photorefractive materials and their applications}, volume 114.
\newblock Springer, 2007.

\bibitem{coufal2000holographic}
Hans~J Coufal, Demetri Psaltis, and Glenn~T Sincerbox.
\newblock {\em Holographic data storage}, volume~8.
\newblock Springer, 2000.

\bibitem{wakayama2013mode}
Yuta Wakayama, Atsushi Okamoto, Kento Kawabata, Akihisa Tomita, and Kunihiro
  Sato.
\newblock Mode demultiplexer using angularly multiplexed volume holograms.
\newblock {\em Optics Express}, 21(10):12920--12933, 2013.

\bibitem{gerke2010aperiodic}
Tim~D Gerke and Rafael Piestun.
\newblock Aperiodic volume optics.
\newblock {\em Nature Photonics}, 4(3):188--193, 2010.

\bibitem{Douglass:22}
Glen Douglass, W.~Minster Kunkel, Ali Ghoreyshi, Simon Gross, Michael~J.
  Withford, and James~R. Leger.
\newblock Two dimensional gradient-index beam shapers fabricated using
  ultrafast laser inscription.
\newblock {\em Optics Express}, 30(22):40592--40598, Oct 2022.

\bibitem{barre2021tomographic}
Nicolas Barr{\'e}, Ravi Shivaraman, Lisa Ackermann, Simon Moser, Michael
  Schmidt, Patrick Salter, Martin Booth, and Alexander Jesacher.
\newblock Tomographic refractive index profiling of direct laser written
  waveguides.
\newblock {\em Optics Express}, 29(22):35414--35425, 2021.

\bibitem{butaite2022build}
Un{\.e}~G B{\=u}tait{\.e}, Hlib Kupianskyi, Tom{\'a}{\v{s}}
  {\v{C}}i{\v{z}}m{\'a}r, and David~B Phillips.
\newblock How to build the optical inverse of a multimode fibre.
\newblock {\em arXiv preprint arXiv:2204.02865}, 2022.

\bibitem{Wright:22}
Logan~G. Wright, William~H. Renninger, Demetri~N. Christodoulides, and Frank~W.
  Wise.
\newblock Nonlinear multimode photonics: nonlinear optics with many degrees of
  freedom.
\newblock {\em Optica}, 9(7):824--841, Jul 2022.

\bibitem{lib2022quantum}
Ohad Lib and Yaron Bromberg.
\newblock Quantum light in complex media and its applications.
\newblock {\em Nature Physics}, 18(9):986--993, 2022.

\bibitem{berlich2016fabrication}
Ren{\'e} Berlich, Daniel Richter, Martin Richardson, and Stefan Nolte.
\newblock Fabrication of computer-generated holograms using femtosecond laser
  direct writing.
\newblock {\em Optics Letters}, 41(8):1752--1755, 2016.

\bibitem{feit1978light}
MD~Feit and JA~Fleck.
\newblock Light propagation in graded-index optical fibers.
\newblock {\em Applied Optics}, 17(24):3990--3998, 1978.

\bibitem{jesacher2010parallel}
Alexander Jesacher and Martin~J Booth.
\newblock Parallel direct laser writing in three dimensions with spatially
  dependent aberration correction.
\newblock {\em Optics Express}, 18(20):21090--21099, 2010.

\bibitem{fontaine2019laguerre}
Nicolas~K Fontaine, Roland Ryf, Haoshuo Chen, David~T Neilson, Kwangwoong Kim,
  and Joel Carpenter.
\newblock Laguerre-gaussian mode sorter.
\newblock {\em Nature Communications}, 10(1):1--7, 2019.

\bibitem{barre2022inverse}
Nicolas Barr{\'e} and Alexander Jesacher.
\newblock Inverse design of gradient-index volume multimode converters.
\newblock {\em Optics Express}, 30(7):10573--10587, 2022.

\bibitem{colburn1971volume}
WS~Colburn and KA~Haines.
\newblock Volume hologram formation in photopolymer materials.
\newblock {\em Applied Optics}, 10(7):1636--1641, 1971.

\bibitem{jialing2019review}
Jian Jialing, Cao Lin, Wei Xiqiao, Guo Jinxin, Wang Dayong, and Zhang Xinping.
\newblock A review of photopolymers on holography volume data storage.
\newblock {\em Opto-Electronic Engineering}, 46(3):180552--1, 2019.

\bibitem{hesselink1998photorefractive}
Lambertus Hesselink, Sergei~S Orlov, Alice Liu, Annapoorna Akella, David Lande,
  and Ratnakar~R Neurgaonkar.
\newblock Photorefractive materials for nonvolatile volume holographic data
  storage.
\newblock {\em Science}, 282(5391):1089--1094, 1998.

\bibitem{yu2014flat}
Nanfang Yu and Federico Capasso.
\newblock Flat optics with designer metasurfaces.
\newblock {\em Nature materials}, 13(2):139--150, 2014.

\bibitem{walther2012spatial}
Benny Walther, Christian Helgert, Carsten Rockstuhl, Frank Setzpfandt, Falk
  Eilenberger, Ernst-Bernhard Kley, Falk Lederer, Andreas T{\"u}nnermann, and
  Thomas Pertsch.
\newblock Spatial and spectral light shaping with metamaterials.
\newblock {\em Advanced Materials}, 24(47):6300--6304, 2012.

\bibitem{scheuer2017metasurfaces}
Jacob Scheuer.
\newblock Metasurfaces-based holography and beam shaping: engineering the phase
  profile of light.
\newblock {\em Nanophotonics}, 6(1):137--152, 2017.

\bibitem{dorrah2022tunable}
Ahmed~H Dorrah and Federico Capasso.
\newblock Tunable structured light with flat optics.
\newblock {\em Science}, 376(6591):eabi6860, 2022.

\bibitem{jang2021independent}
Junhyeok Jang, Gun-Yeal Lee, Jangwoon Sung, and Byoungho Lee.
\newblock Independent multichannel wavefront modulation for angle multiplexed
  meta-holograms.
\newblock {\em Advanced Optical Materials}, 9(17):2100678, 2021.

\bibitem{oh2022adjoint}
Jaewon Oh, Kangmei Li, Jun Yang, Wei~Ting Chen, Ming-Jun Li, Paulo Dainese, and
  Federico Capasso.
\newblock Adjoint-optimized metasurfaces for compact mode-division
  multiplexing.
\newblock {\em ACS Photonics}, 9(3):929--937, 2022.

\bibitem{jalali2006silicon}
Bahram Jalali and Sasan Fathpour.
\newblock Silicon photonics.
\newblock {\em Journal of Lightwave Technology}, 24(12):4600--4615, 2006.

\bibitem{siew2021review}
S.~Y. Siew, B.~Li, F.~Gao, H.~Y. Zheng, W.~Zhang, P.~Guo, S.~W. Xie, A.~Song,
  B.~Dong, L.~W. Luo, C.~Li, X.~Luo, and G.-Q. Lo.
\newblock Review of silicon photonics technology and platform development.
\newblock {\em Journal of Lightwave Technology}, 39(13):4374--4389, 2021.

\bibitem{trinh1997silicon}
PD~Trinh, S~Yegnanarayanan, F~Coppinger, and B~Jalali.
\newblock Silicon-on-insulator (soi) phased-array wavelength
  multi/demultiplexer with extremely low-polarization sensitivity.
\newblock {\em IEEE Photonics Technology Letters}, 9(7):940--942, 1997.

\bibitem{piggott2015inverse}
Alexander~Y Piggott, Jesse Lu, Konstantinos~G Lagoudakis, Jan Petykiewicz,
  Thomas~M Babinec, and Jelena Vu{\v{c}}kovi{\'c}.
\newblock Inverse design and demonstration of a compact and broadband on-chip
  wavelength demultiplexer.
\newblock {\em Nature Photonics}, 9(6):374--377, 2015.

\bibitem{li2019multimode}
Chenlei Li, Dajian Liu, and Daoxin Dai.
\newblock Multimode silicon photonics.
\newblock {\em Nanophotonics}, 8(2):227--247, 2019.

\bibitem{minzioni2017roadmap}
Paolo Minzioni, Roberto Osellame, Cinzia Sada, S~Zhao, F~G Omenetto, Kristinn~B
  Gylfason, Tommy Haraldsson, Yibo Zhang, Aydogan Ozcan, Adam Wax, Frieder
  Mugele, Holger Schmidt, Genni Testa, Romeo Bernini, Jochen Guck, Carlo
  Liberale, Kirstine Berg-S{\o}rensen, Jian Chen, Markus Pollnau, Sha Xiong,
  Ai-Qun Liu, Chia-Chann Shiue, Shih-Kang Fan, David Erickson, and David
  Sinton.
\newblock Roadmap for optofluidics.
\newblock {\em Journal of Optics}, 19(9):093003, 2017.

\bibitem{photonic_packaging}
Lee Carroll, Jun-Su Lee, Carmelo Scarcella, Kamil Gradkowski, Matthieu
  Duperron, Huihui Lu, Yan Zhao, Cormac Eason, Padraic Morrissey, Marc Rensing,
  Sean Collins, How~Yuan Hwang, and Peter O’Brien.
\newblock Photonic packaging: Transforming silicon photonic integrated circuits
  into photonic devices.
\newblock {\em Applied Sciences}, 6(12), 2016.

\bibitem{Yoo2016}
S.~J.~Ben Yoo, Binbin Guan, and Ryan~P. Scott.
\newblock Heterogeneous 2d/3d photonic integrated microsystems.
\newblock {\em Microsystems {\&} Nanoengineering}, 2(1):16030, Aug 2016.

\bibitem{strickler1991three}
James~H Strickler and Watt~W Webb.
\newblock Three-dimensional optical data storage in refractive media by
  two-photon point excitation.
\newblock {\em Optics Letters}, 16(22):1780--1782, 1991.

\bibitem{glezer1997three}
EN~Glezer, M~Milosavljevic, L~Huang, RJ~Finlay, T-H Her, JP~Callan, and
  E~Mazur.
\newblock Three-dimensional optical storage inside transparent materials:
  errata.
\newblock {\em Optics Letters}, 22(6):422--422, 1997.

\bibitem{zhang2014seemingly}
Jingyu Zhang, Mindaugas Gecevi{\v{c}}ius, Martynas Beresna, and Peter~G
  Kazansky.
\newblock Seemingly unlimited lifetime data storage in nanostructured glass.
\newblock {\em Physical Review Letters}, 112(3):033901, 2014.

\bibitem{xu2022femtosecond}
Xiaoyi Xu, Tianxin Wang, Pengcheng Chen, Chao Zhou, Jianan Ma, Dunzhao Wei,
  Huijun Wang, Ben Niu, Xinyuan Fang, Di~Wu, Shining Zhu, Min Gu, Min Xiao, and
  Yong Zhang.
\newblock Femtosecond laser writing of lithium niobate ferroelectric
  nanodomains.
\newblock {\em Nature}, 609(7927):496--501, Sep 2022.

\bibitem{dellavalle2008}
Giuseppe Della~Valle, Roberto Osellame, and Paolo Laporta.
\newblock Micromachining of photonic devices by femtosecond laser pulses.
\newblock {\em Journal of Optics A: Pure and Applied Optics}, 11(1):013001,
  2008.

\bibitem{bisch2019}
Niklas Bisch, Jun Guan, Martin~J Booth, and Patrick~S Salter.
\newblock Adaptive optics aberration correction for deep direct laser written
  waveguides in the heating regime.
\newblock {\em Applied Physics A}, 125:364, 2019.

\bibitem{fienup1993phase}
James~R Fienup.
\newblock Phase-retrieval algorithms for a complicated optical system.
\newblock {\em Applied Optics}, 32(10):1737--1746, 1993.

\end{thebibliography}

\clearpage

\appendix

\section{Appendix}

\section{APVE characterization}\label{sec:APVE_carac}

\begin{figure}[!b]
  \centering
  \includegraphics[width=\textwidth]{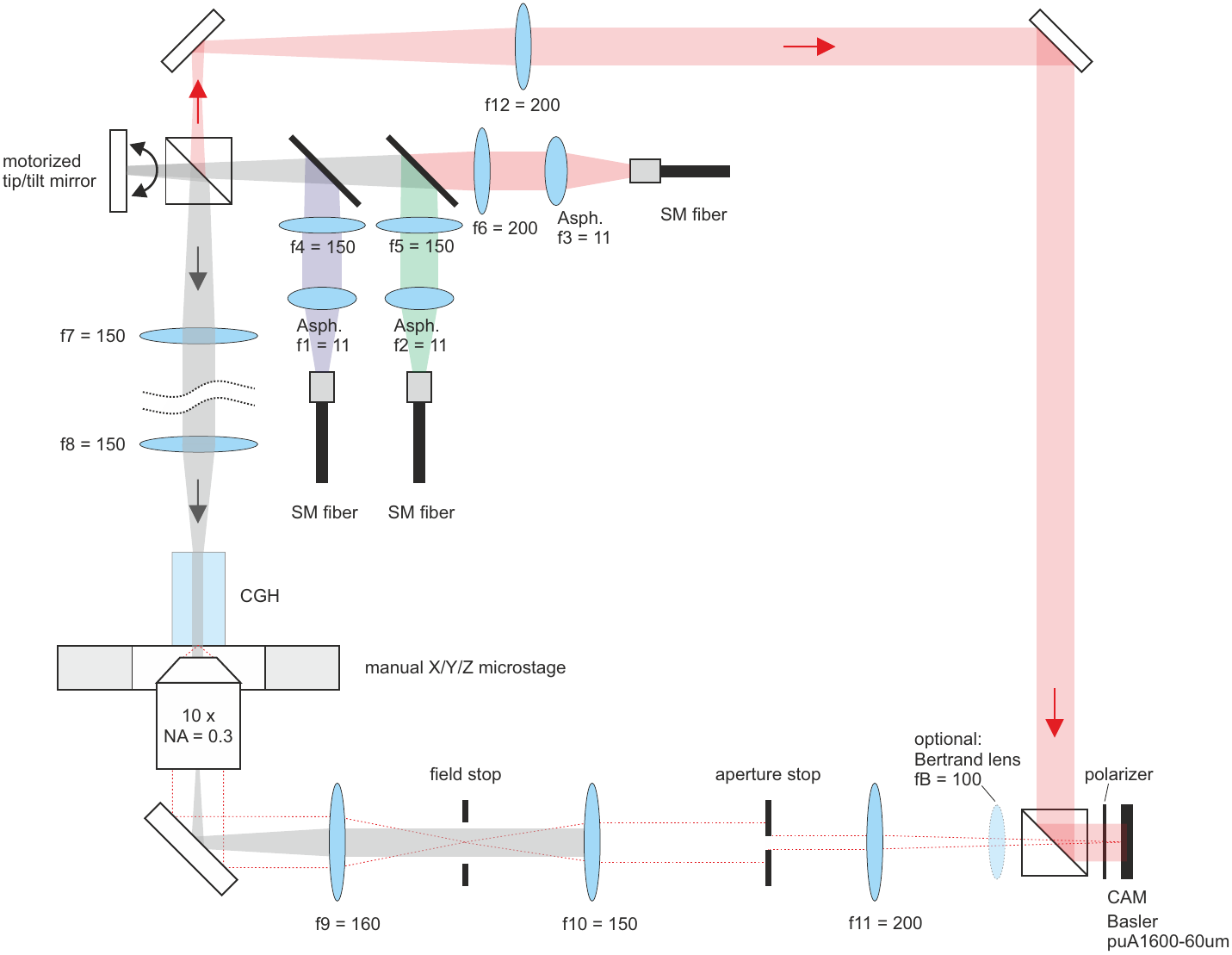}
  \caption{\textbf{Optical setup for characterizing APVEs.} A
    motorized tip/tilt mirror is imaged onto the input facet of the
    APVE, which rests on a manual x/y microstage. The APVE can be
    simultaneously exposed to three laser beams (red/green/blue) or
    optionally to light form the monochromator (not shown). The
    APVE output facet is imaged from below, where the imaging NA
    can be controlled by an aperture stop. The setup further
    features a Mach-Zehnder interferometer for characterizing the
    mode sorter APVE.}
  \label{fig:setup}
\end{figure}
  
The APVEs are characterized with the home-built imaging system shown
in Fig.~\ref{fig:setup}. The setup allows one to focus fiber coupled
laser beams of three different wavelengths (Toptica ibeam smart
@640~nm, Lasos LGK7786P @543~nm, Thorlabs laser diode PL450B @455~nm)
onto the input facet of a APVE, which rests on a manual 3D micro
stage. The waists of the laser foci can be controlled by choosing
appropriate focal lengths for the lenses 4, 5 and 6. A motorized
tip/tilt mirror is imaged onto the APVE input facet using a Keplerian
telescope. This allows one to precisely control the laser's AOI.  The
output facet is imaged using a microscope objective lens (Olympus
UPlanFL, 10x, 0.3 NA), whose exit pupil is imaged onto an iris with
tunable aperture. This iris was used to restrict the imaging NA to
0.05 when characterizing the mode sorter. Finally, an image of the
end facet is formed at the sensor of a CCD camera (Basler
puA1600-60uA). Optionally, a Bertrand lens can be flipped in for
imaging the objective's exit pupil, which is used to calibrate the
laser AOIs.

The setup further comprises a Mach-Zehnder interferometer for
measuring complex field distributions at the output of the mode sorter
APVE using off-axis interferometry.

\subsection*{Interferometry}
\begin{figure}[!ht]
    \centering
    \includegraphics[width=0.8\textwidth]{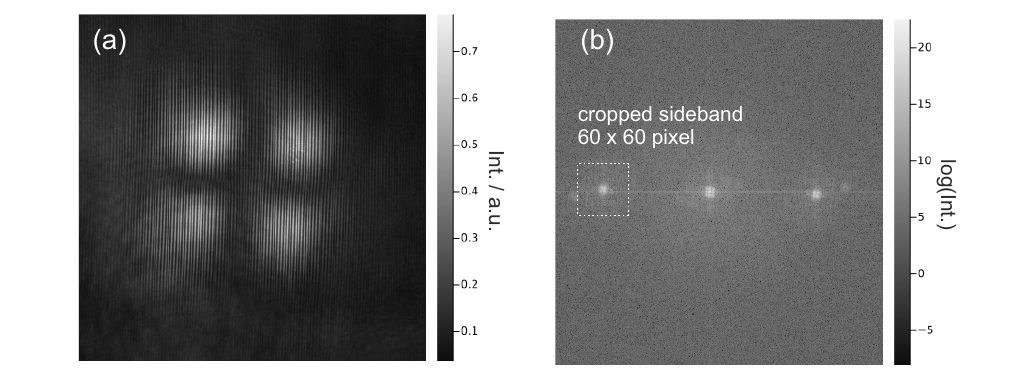}
    \caption{\textbf{Interferometry to characterize the light field
        shaped by the mode sorter.} Left: Raw interferogram captured
      by the camera; Right: Its logarithmic power spectrum. One
      sideband is cropped for further processing.}
    \label{fig_sup_interferometry}
\end{figure}

To characterize the quality of the mode sorter, it is required to
measure the complex light field at the output facet of the APVE. This
is done using off-axis interferometry using a Mach-Zehnder
interferometer. The reference wave is tilted, such that the
interference fringes have period of about 3 pixels as shown in
Fig.~\ref{fig_sup_interferometry}~(a). The intensity image is Fast
Fourier transformed (FFT), followed by cropping a 60$\times$60 pixel
region around one sideband as shown in
Fig.~\ref{fig_sup_interferometry}~(b).  The complex field is then
obtained by taking an inverse FFT of the cropped sideband. However, at
this stage it will most likely show only a poor overlap with any HG
mode. This is because of several factors: Firstly, the field is
usually not perfectly centered on the pixel grid and there could be a
small geometric rotation in the experimental images. Secondly, the
presence of a global phase tilt and defocus is likewise reducing the
overlap integral.  Finally, the waist of the measured mode could be
slightly different compared to the simulation.

To solve these issues, we employed a Nelder Mead optimizer (Julia
programming language, toolbox: Optim.jl) to find those seven
parameters which maximize the overlap integral between light field
obtained at 0$^\circ$ AOI and the matching target mode
$\mathrm{HG}_{1,1}$.  These parameters are then used to correct all
experimentally recorded fields.

\section{Additional information on the colour multiplexer}\label{sec:Additional_color_multiplexer}

\begin{figure}[!ht]
  \centering \includegraphics[width = 11cm]{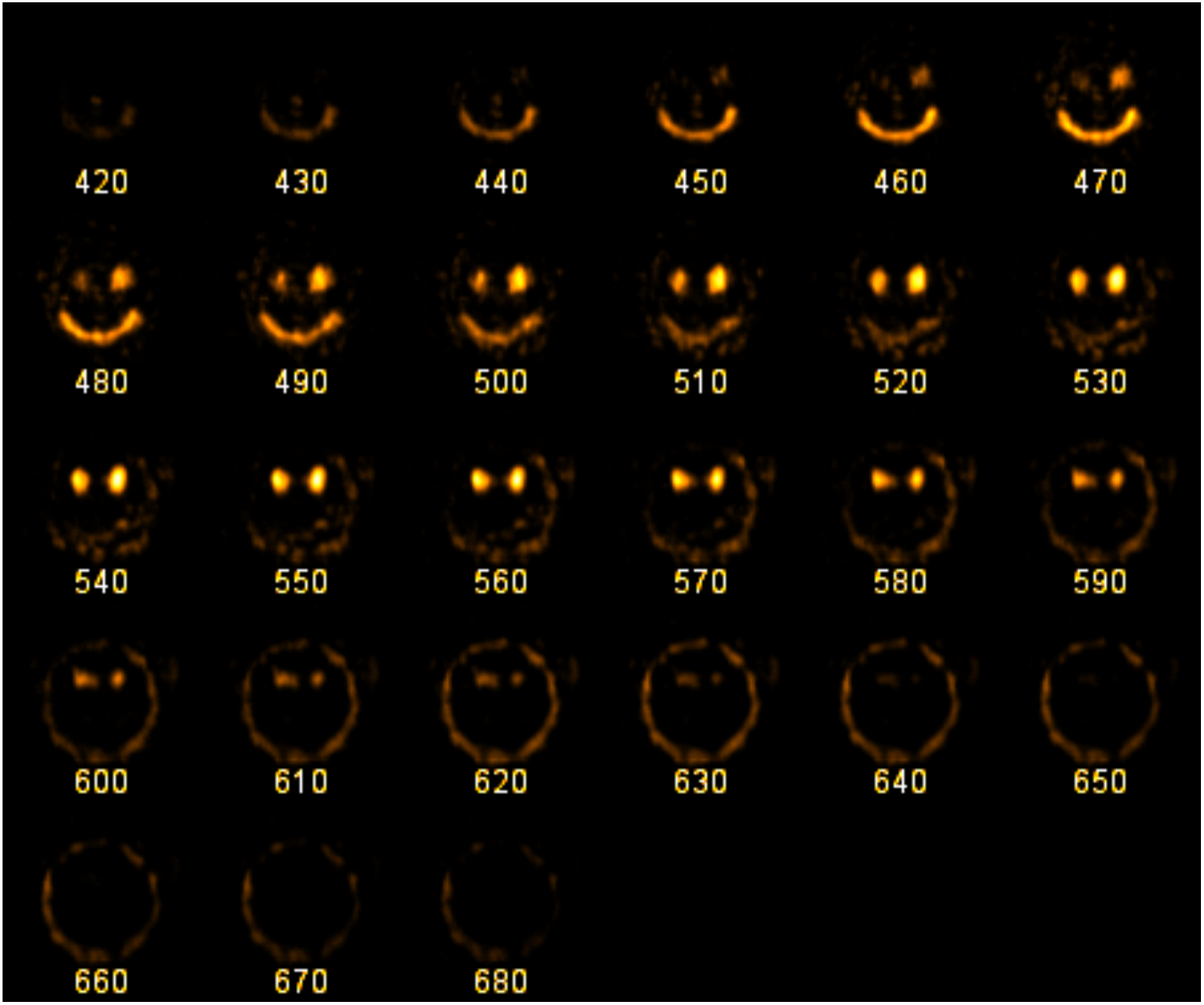}
  \caption{\textbf{Wavelength dependent output intensities of the
      colour smiley APVE.} The illumination light was provided by
    a tunable monochromator. The numbers below the images denote the
    respective wavelength in nm.}
    \label{fig:montage monochromator}
\end{figure}

Figure~\ref{fig:montage monochromator} shows output intensities of the
colour smiley APVE, taken with the monochromator light source at
wavelengths varied from 420~nm to 680~nm in steps of 10~nm.  For APVE
characterization, the light form the monochromator's arc lamp was
focused into a light guide of 1 mm diameter. About 1~cm from light
guide's output we placed a 100~\textmu m pinhole to further increase
the spatial coherence. This pinhole was then imaged with a slight
demagnification (telescope with 200 mm and 150 mm lenses) onto the
motorized tip/tilt mirror shown in Fig.~\ref{fig:setup}, such that the
monochromator light at the input facet of the APVEs took the form of a
Gaussian with $w_0 \approx 30$~\textmu m.

\section{Additional information on the mode sorter}\label{sec:Additional_mode_sorter}

\begin{figure}[!ht]
    \centering
    \includegraphics[width=0.8\textwidth]{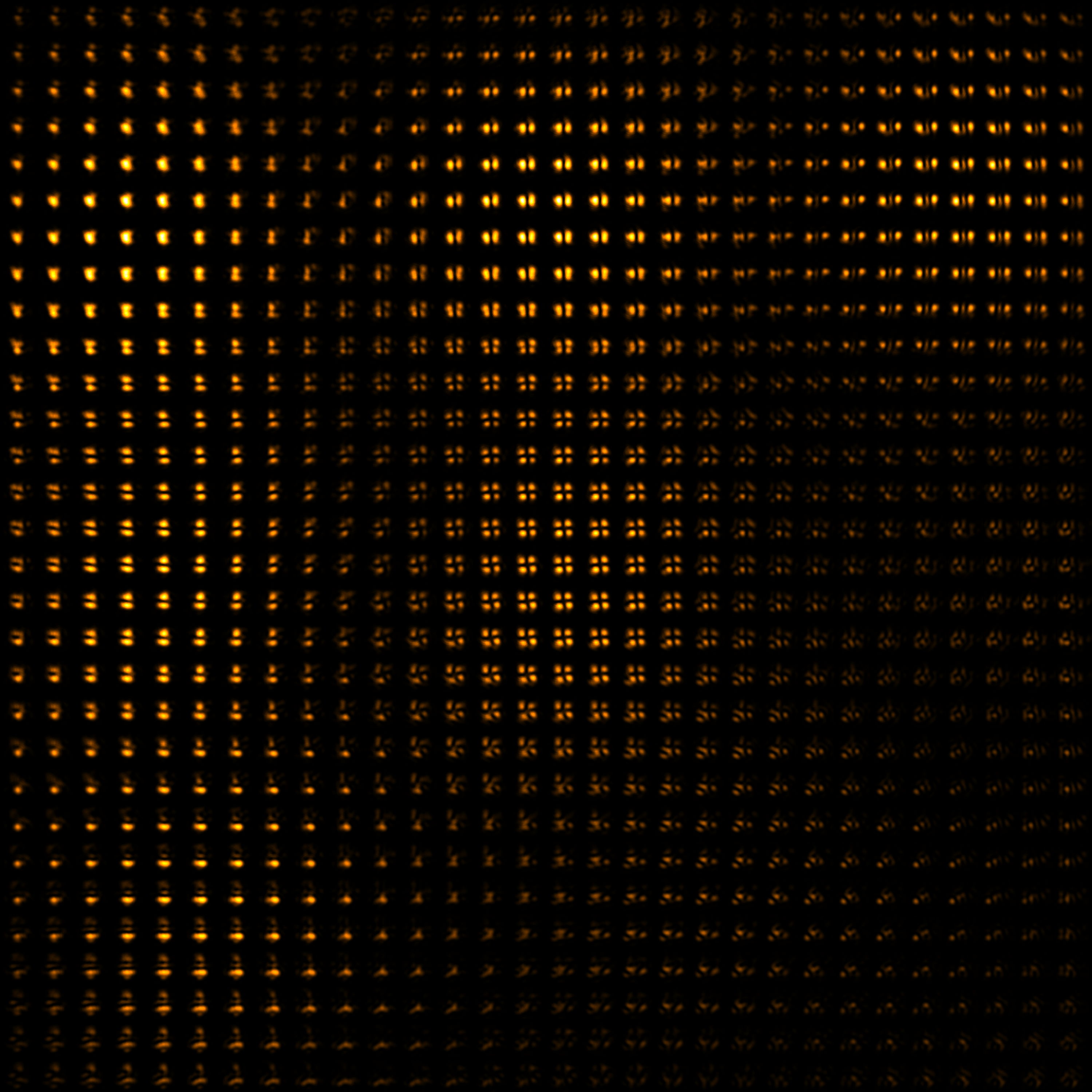}
    \caption{\textbf{Angular dependent readouts of the mode sorter
        APVE.} The figure shows 30$\times$30 images, each one showing
      the intensity at the APVE output facet for a different
      incidence angle of the readout beam. The angular increment from
      one image to the next is 0.11 degrees. The image showing the
      $HG_{1,1}$ mode in the center corresponds to a readout angle of
      zero.}
    \label{fig:intensity stack}
\end{figure}

\begin{figure}[!ht]
  \centering \includegraphics[width=0.6\textwidth]{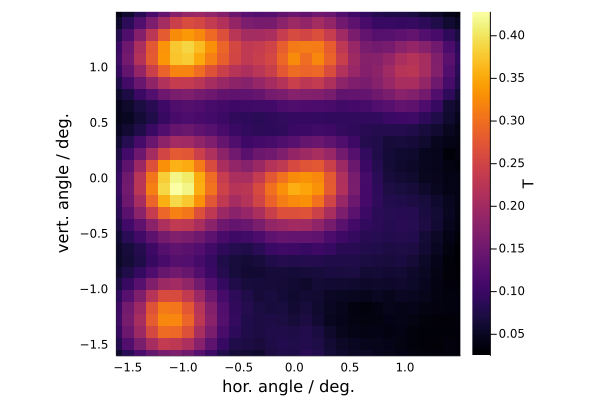}
  \caption{\textbf{Angular dependence of the transmission $T$ of the
      mode sorter.} The figure shows the measured power fraction of
    the input beam arriving the camera sensor.}
    \label{fig:angular transmission}
\end{figure}

\begin{figure}[!ht]
  \centering \includegraphics[width=0.8\textwidth]{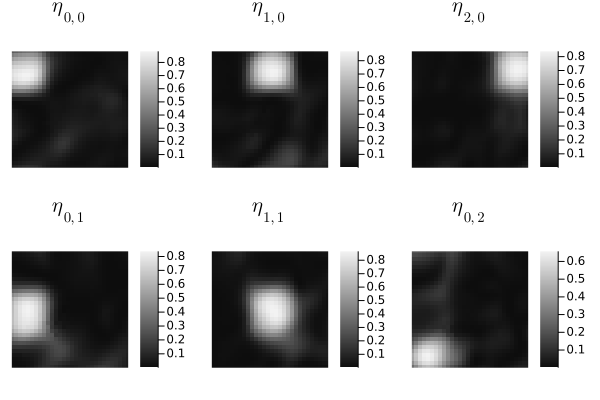}
  \caption{\textbf{Angular dependence of the experimental efficiency
      values $\eta_{i,j}$ of the mode sorter.}}
    \label{fig:angular OI}
\end{figure}

Figure~\ref{fig:intensity stack} presents a table of 30$\times$30
intensity images, which appear at the output facet of the mode sorter
when the AOI of the input beam is altered in increments of 0.11
degrees (in air). One can clearly identify six angular regions where
the output pattern resembles a particular HG mode.
Figure~\ref{fig:angular transmission} shows the measured transmission
$T$ of the mode sorter as a function of the laser AOI. One can clearly
define six local maxima, whose respective peak values match the
numbers in table~\ref{tab:transmissions sorter}. Finally,
Fig.~\ref{fig:angular OI} visualizes the measured angular dependence
of all six efficiency values $\eta_{i,j}$ as defined by
Eq.~\ref{eq:OI}.

\clearpage

\section{Manufacturing of APVEs}\label{sec:Manufacturing_APVEs}

The APVEs were manufactured using the direct laser writing
technique~\cite{dellavalle2008}. The beam from an ultrafast laser
(Light Conversion, Pharos SP 6W) emitting pulses with 170~fs duration,
at a repetition rate of 1~MHz and wavelength 515~nm was expanded and
directed toward a phase-only liquid crystal spatial light modulator
(SLM) (Hamamatsu X10468). The SLM was imaged in 4f configuration onto
the pupil plane of a microscope objective (Zeiss 20$\times$
0.5NA). The beam was thus focused into the glass sample (Corning EAGLE
2000) to fabricate the APVE voxels. The sample was mounted on a
precision 3D translation stage (Aerotech ABL10100 (x-z), ANT-95V (y))
and moved through the laser focus at a speed of 2~mm/s in the $z$
direction, as defined in Figure~\ref{fig:concept}~(a). The PSO function
was used to gate the laser according to the design specification and
fabricate voxels at the desired position. The laser pulse energy was
37~nJ and the polarisation was linear along the translation
direction. We note that this pulse energy is significantly lower than
that which would typically be used to write single mode optical
waveguides in such a system. As the sample was moved in the $y$
direction to build up different layers of the APVE, the phase pattern
displayed on the SLM was updated to compensate for depth-induced
spherical aberration arising from refraction at the sample
surface~\cite{bisch2019}.

\section{Design algorithm}\label{sec:Design_algorithm}

In the following, we present an algorithm that allows to design
voxel-based APVEs for the conversion of $N$ mutually incoherent
transverse input modes $\{u_n\}_{1\leq n \leq N}$ to $N$ target modes
$\{v_n\}_{1\leq n \leq N}$, with a one-to-one mapping. In general, any
pair of modes $(u_n,v_n)$ has its own wavelength $\lambda_n$.  We
consider the design of a finite extent sample in a rectangular
coordinate system ($O,x,y,z$), $z$ being the direction of propagation.
We denote $z_{in}$ the transverse plane where the input modes $\{u_n\}$ are defined,
and $z_{out}$ the destination plane where the target modes $\{v_n\}$ are
defined. The bulk refractive index of the sample at wavelength
$\lambda_n$ is denoted $n_n$. In terms of fabrication capabilities, we
assume that we have access to a diversity of transverse voxel
distributions that we are able to write in the glass sample at any 3D
location, and that the length of any voxel (i.e., the $z$ dimension)
can be freely chosen. One example of a voxel cross section is given in
Figure~\ref{fig:concept}~(b), but many more are possible by varying
the writing parameters such as the pulse energy or the aberration
compensation term. Let us note $D$ the number of different transverse
voxel distributions $\{\Delta\mathrm{RI}_{d}\}_{1 \leq d \leq D}$
which we consider to use for the inverse design. These transverse
distributions are relative refractive index changes compared to the
bulk refractive index, and are possibly wavelength dependent. By
convention, we also define $\Delta\textrm{RI}_0$ to refer to a null
refractive index modification. In order to make the inverse design
much simpler, the sample volume is virtually split into sub-volumes
$\varv_{ijk}$, $1 \leq i \leq I$, $1 \leq j \leq J$,
$1 \leq k \leq K$, acting as placeholders for the elementary voxel
distributions $\{\Delta\mathrm{RI}_{d}^{ijk}\}$. These sub-volumes
extensions in 3 dimensions are denoted $\mathrm{D}x$, $\mathrm{D}y$
and $\mathrm{D}z$. Thus, the possible lengths of the waveguides used
for the design can only be integer multiples of the elementary length
$\mathrm{D}z$. In our designs, this elementary length is typically
$\mathrm{D}z = 10\,\textrm{\textmu m}$. For the simulation to be numerically
accurate, the sampling resolution in 3 dimensions $\mathrm{d}x$,
$\mathrm{d}y$ and $\mathrm{d}z$ must be much better, typically
$\mathrm{d}x = \mathrm{d}y = 0.25\,\textrm{\textmu m}$ and
$\mathrm{d}z = 0.5\,\textrm{\textmu m}$.

\paragraph{Forward model\newline}
The propagation of the input modes $\{u_n\}$ from the plane $z_{in}$
to the plane $z_{out}$ relies on the split-step Fourier beam
propagation method (BPM), with spatial resolutions $\mathrm{d}x$,
$\mathrm{d}y$ and $\mathrm{d}z$. The input modes cross the sub-volumes
$\varv_{ijk}$ layer by layer, for $k$ varying from 1 to $K$. Each
sub-volume $\varv_{ijk}$ may contain (or not) one of the possible
transverse distributions $\{\Delta\textrm{RI}_d^{ijk}\}$, depending on
the choice that was made in a previous iteration by the selection
algorithm that we present next. During this forward propagation, the
transverse distributions $\{u_n\}$ are saved before crossing each
sub-volume $\varv_{ijk}$, and we note them $\{u^{ijk}_n\}$ for future
reference. We also note $\{z_k\}$ the common transverse planes where
these distributions are saved.

According to the BPM, for a given input mode $u_n$, crossing each
layer of thickness $\mathrm{d}z$ is modeled as a small propagation of
length $\mathrm{d}z$ through an homogeneous medium of refractive index
$n_n$, followed by the multiplication of a local phase $\varphi_{ijk}$
depending on the activated region:
\begin{equation}
  \varphi_{ijk}(\lambda_n) = \exp\left(i\frac{2\pi}{\lambda_n}\Delta\textrm{RI}^{ijk}_d(\lambda_n)\mathrm{d}z\right),\,0\leq d \leq D.
\end{equation}

\paragraph{Error metric\newline}

Our inverse design relies on a multi-objective minimization based on
the squared $l_2$ distance between the input and target modes:
\begin{equation}
  O_{min} = \min\left\{\sum\limits_{n=1}^N\|u_n-v_n\|_2^2\right\} = \min\left\{\sum\limits_{n=1}^N{\int\limits_{\,\,z_{out}}\vert u_n - v_n \vert^2}\right\}.
  \label{eq:min_error}
\end{equation}

We can expand the squared modulus as $\vert u_n - v_n \vert^2 = \vert
u_n\vert^2 + \vert v_n\vert^2 - 2 \Re{\left(v_n^*u_n\right)}$, which
allows to turn the minimization problem~(\ref{eq:min_error}) into the
following maximization problem in the case of a lossless design
($\|\|_2^2$ of $u_n$ and $v_n$ is conserved):

\begin{equation} O_{max} =
\max\limits_{\left\{\Delta\mathrm{RI}_d^{ijk}\right\}}\left\{\sum\limits_{n=1}^N{\int\limits_{\,\,z_{out}}\Re\left(v_n^*u_n\right)}\right\}.
    \label{eq:max_error}
\end{equation}

The expression~(\ref{eq:max_error}) consists only of overlap integrals
that are conserved during propagation (or backpropagation) through a
lossless optical system, thus allowing for an iterative propagation
algorithm (IPA)~\cite{fienup1993phase}, where the same criterion
$O_{max}$ is iteratively maximized in each plane $z_k$ (in a reverse
fashion).

\paragraph{Voxel distribution selection algorithm\newline}

We assume that we know the transverse distributions $\{v_n^{ijk+1}\}$
corresponding to the local distributions of the target modes $\{v_n\}$
in the plane $z_{k+1}$ just after the sub-volumes $\{\varv_{ijk}\}$,
and the transverse distributions $\{u_n^{ijk}\}$ of the input modes
$\{u_n\}$ stored just before the same sub-volumes. The selection of
the refractive index distribution $\Delta\textrm{RI}_d^{ijk}$ relies
on the maximization of the local multi-objective criterion
\begin{equation} O_{max}^{ijk} =
\max\limits_{d}\left\{\sum\limits_{n=1}^N{\int\limits_{\,\,z_{k+1}}\Re\left((v_n^{ijk+1})^*u_n^{ijk+1}(d)\right)}\right\},
    \label{eq:max_local}
\end{equation} where $u_n^{ijk+1}(d)$ denotes the local input modes
distribution that have been propagated from plane $z_k$, where they
were stored during the forward pass, to plane $z_{k+1}$ through a
local refractive index distribution $\Delta\textrm{RI}_d^{ijk}$. In
order to find $d$ maximizing $O_{max}^{ijk}$, all the distributions
$\{\Delta\textrm{RI}_d^{ijk}\}$ need to be tried for each sub-volume
$\varv_{ijk}$ ($k$ fixed).

Once the $\{\Delta\textrm{RI}_d^{ijk}\}$ distributions have been
selected for a given $k$, the target modes $\{v_n^{ijk+1}\}$ are
backpropagated to the plane $z_{k}$, and the same procedure is
repeated iteratively until the selection of the
$\{\Delta\textrm{RI}_d^{ij1}\}$.

\paragraph{Iterative propagation algorithm\newline}

We summarize the different steps of the optimization procedure below:

\begin{enumerate}
    \item Propagate the input modes $\{u_n\}$ from their definition
plane $z_{in} = z_1$ to the destination plane $z_{out} = z_{K+1}$
using BPM, taking into account the current waveguide selection in each
sub-volume $\varv_{ijk}$ ($1 \leq k \leq K$). At the beginning of the
first iteration, all $\{\varv_{ijk}\}$ contain a null refractive index
distribution ($\Delta\textrm{RI}_0$). During propagation, store the
local field distributions $\{u_n^{ijk}\}$ at planes $\{z_k\}$.
    \item Compute the final overlap criterion~(\ref{eq:max_error}) for
monitoring the convergence of the algorithm. If the convergence is
satisfactory or stagnant, then exit the optimization procedure and
consider the current waveguide selection as the optimal design.
    \item Iteratively backpropagate the target modes $\{v_n\}$ from
plane $z_{out} = z_{K+1}$ to plane $z_{in} = z_1$, while proceeding to
a new voxel selection in each sub-volume $\varv_{ijk}$ according to
criterion~(\ref{eq:max_local}). For a given $k$, this criterion leads
to the best voxel selection in each sub-volume, after which the modes
$\{v_n\}$ in plane $z_{k+1}$ can be backpropagated to plane $z_{k}$,
and this iteratively until $k=1$.
    \item Repeat the procedure from step 1.
\end{enumerate}

\end{document}